\documentclass[%
 reprint,
 amsmath,amssymb,
 aps,showkeys
]{revtex4-1}

\usepackage{graphicx}
\usepackage{dcolumn}
\usepackage{bm}
\usepackage{hyperref}
\usepackage{float}

\begin{document}

\preprint{APS/123-QED}

\title{Quantum theory of the plasmonic nanolaser}

\author{I. Fyodorov}
\email{ilfedorov@gmail.com}
\affiliation{Moscow Institute of Physics and Technology, Moscow, Russia}

\author{A. Sarychev}%
\email{sarychev_andrey@yahoo.com}

\affiliation{Institute for Theoretical and Applied Electromagnetics RAS, Moscow, Russia}

\date{\today}

\begin{abstract}
In this work, we present a fully quantum theory of the plasmonic nanolaser,
based on the maser model. Theory can be applied both to the microlasers with
high Q-factor cavities and plasmonic nanolasers. We show that the latter is
essentially a thresholdless device. We obtain the statistics of quanta,
spectrum of the plasmonic radiation and the second-order coherence degree in
the steady state. The limits of the model applicability are discussed in
detail. All results are compared with the full numerical simulation.

\end{abstract}

\keywords{plasmonic nanolaser, spaser, luminescence, quantum theory}

\maketitle


\section{\label{sec:intro}Introduction}

The modern story of the utilization of the nanoparticle-living plasmons started a decade ago, when the possibility of the enhancement of surface plasmons in the nanoparticles (NPs) by optical gain in dielectric medium was predicted theoretically \cite{Bergman2003} and demonstrated in the experiments \cite{Noginov2006, Noginov2006a}.

It has been realized, that extremely tight field confinement in surface plasmon modes can be used to achieve the strong coupling regime with the emitters \cite{Chang2007}. The metal NPs became a promising platform for implementation of the great variety of effects known in quantum optics and cavity QED \cite{Chang2006, Kimble1998, Akimov2007, Dominguez2013}.
In parallel, papers \cite{Noginov2009, Oulton2009} claimed the lasing on the individual NPs, and a lot of work focused on the realization of smaller and faster sources of light \cite{Nomura2010, Nezhad2010, Ma2010, Wu2011, Lu2012, Suh2012}.

A metal NP adjacent to the active medium appeared as a basic unit cell of many applications of nanoplasmonics, such as metamaterials, nanosensing, optical logic, etc. Though, theoretical description of this object (which essentially constitutes a nanolaser, alternatively known as spaser) trails far behind the experimental progress. In ref.\cite{Bogdanov2013} we showed that semiclassical limit of quantum theory (see, e.g. \cite{Protsenko2005, Stockman2010, Andrianov2011}), which is eventually equivalent to the classical Maxwell-Bloch approach \cite{Sarychev2007}, is unadaptable for the nanolaser. It corresponds to the so-called thermodynamic limit of the laser \cite{Rice1994}, which is not the case, since in the nano-sized resonant cavity most of the spontaneous radiation of the emitters goes directly to the laser mode.

Several works, addressing operation of the small lasers, go beyond the classical approximation. Papers \cite{Mu1992, Benson1999} are of great help for understanding the single-emitter laser operation. Unfortunately, these results cannot be directly transferred to the case of nanolaser, since the latter has relatively poor Q-factor and needs $ \approx 10^2-10^3 $ \cite{Noginov2009, Suh2012} emitters per NP to achieve sufficient gain. Treating of the problem in Heizenberg picture is performed in \cite{Protsenko1999}. This work perfectly illustrates where the features of quantum behavior should be neglected to obtain a system of equations, which is solvable analytically. In \cite{Bogdanov2013}, we have shown that the number of quanta populating the plasmon oscillator cannot exceed the ceiling, which is of order of tens, due to thermal limitations. That is, the nanolaser is truly a nonlinear noise device, where fluctuations play a crucial role.

Here, for the first time, we make an attempt to describe the plasmonic nanolaser in the formalism of the density matrix. We use a low-loss approximation, which holds for the plasmonic nanolasers with cavities starting from the FOM of $ \gtrsim 10^2 $, which is the case for many realizations \cite{Hill2007, Nezhad2010, Kang2011, Ma2010, Nomura2010}. This allows one to employ the maser approximation and solve the master equation analytically. 

In the second section, we perform quantization of the plasmon a bit neater than it is done in ref.\cite{Bergman2003}. In chapter 3, we describe the model of the active medium and present the master equation. We solve it and derive the applicability conditions for our model. We explore the statistics of quanta and number of plasmons. Fourth chapter is devoted to the coherence properties of the plasmonic radiation. We compare our analytical model with the numeric simulation and discuss the results.

\section{\label{sec:plq}Plasmon quantization}

Since plasmons are bosons, they are expected to be quantized similar to electromagnetic waves. But energy of the plasmon is not just that of the associated electromagnetic mode. Plasmons are electromechanical oscillations, and kinetic energy of electrons is an important contribution. 

\begin{figure}[htbp]
    \includegraphics[width=90mm]{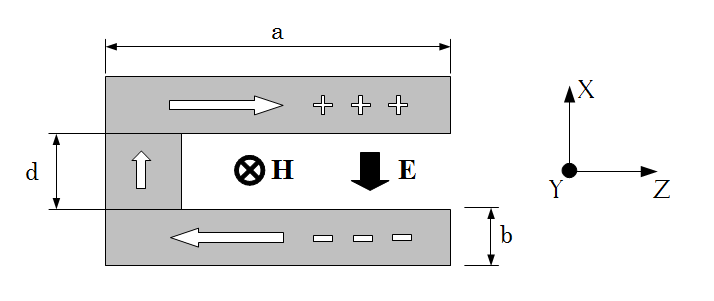}
    \caption{Schematic view of the horseshoe. White rows indicate current direction in the antisymmetric mode. Electric and magnetic fields are parallel to the $ X $ and $ Y $ axes. Depth of the horseshoe in the $ y $ direction is $ l $.}
    \label{p1}
\end{figure}

In \cite{Bergman2003}, plasmon quantization is performed in a declarative manner, that is, without connection with the classical laws of motion. Here, we do the same thing explicitly for the plasmon in the waveguide-like geometry, which behavior is well known (\cite{Sarychev2007}, \cite{Bogdanov2013}). We consider the simple case of the horseshoe geometry, shown in the fig.\ref{p1}.
Charge and current distributions on its arms are found in the form
\begin{equation}
\begin{split}
& J = \sum_{n} J_n \cos(k_n z), \\ 
& Q = \sum_{n} Q_n \sin(k_n z), ~~ k_n=\frac{\pi}{a}\left( n+\frac{1}{2} \right),
\end{split}
\label{eq11}
\end{equation}
where $ J_n $ and $ Q_n $ here are some functions of time. Canonical variables of the system can be chosen as
\begin{equation}
\begin{split}
& p_n = \sqrt{2 \pi V} Q_n, \\ 
& q_n = \dfrac{\sqrt{2 \pi V}}{k_n} \left( \dfrac{1}{c^2} + \dfrac{2}{bd \omega_p^2} \right) J_n,
\end{split}
\label{eq12}
\end{equation}
where $ V=adl $ is cavity volume and $ \omega_p $ is plasma frequency of the metal. The Hamiltonian of n-th mode then reads
\begin{equation}
H_n = \dfrac{1}{2} \left( p_n^2 + \nu_n^2 q_n^2 \right), ~~\nu_n^2 = \omega_p^2 \dfrac{bd k_n^2}{2 + bd \omega_p^2 / c^2}.
\label{eq13}
\end{equation}

It is straightforward to check that the canonical equations of motion with the Hamiltonian \ref{eq13} reproduce the Ohm’s and the charge conservation laws. Now quantization can be performed in the usual way. We let $ q_n $ and $ p_n $ to be operators with the commutation rule $ \left[ q_n, p_n \right] = i $ ($ \hbar $ is taken to be 1 throughout the paper). Annihilation operator of the plasmon then is
\begin{equation}
a_n = \dfrac{1}{\sqrt{2 \nu_n}} \left( \nu_n q_n + i p_n \right), ~~ \left[ a_n, a_n^+ \right] = 1,
\label{eq14}
\end{equation}
and Hamiltonian \ref{eq13} takes the familiar form:
\begin{equation}
H_n = \nu_n \left( a_n^+ a_n + \dfrac{1}{2} \right) .
\label{eq15}
\end{equation}
Operator of the amplitude of the electric field at the arbitrary point inside the cavity is then
\begin{equation}
i \boldsymbol{x} \sqrt{\dfrac{4 \pi \nu}{V}} \sin (k_n z) \left( a_n^+ - a_n \right) ,
\label{eq16}
\end{equation}
where $ \boldsymbol{x} $ is a unit vector in the $ x $ direction.

\section{\label{sec:model}Model and solution}

In this section, we describe and solve our model. Essentially, it is the well-known maser model (see, for example, \cite{Scully1997,Meystre1999}). We shortly introduce it in the next subsection, and then solve it considering features of our problem.

\subsection{\label{sec:mastereq}Master equation}

\begin{figure}[htbp]
    \includegraphics[width=40mm]{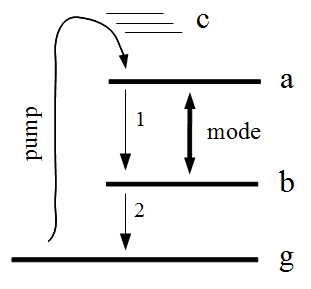}
    \caption{Level scheme of the active unit. Transition between levels $ a $ and $ b $ is resonant with the plasmonic mode. Pumping mechanism brings the system from the ground state $ \left| g \right\rangle  $ to $ \left| a \right\rangle  $ via the short-lived band $ c $.}
    \label{p2}
\end{figure}

For interaction of the plasmonic mode with the active medium we employ the usual laser scheme. We treat every unit of the active medium (atom) as the four-level system (fig.\ref{p2}) which effectively represents both dyes and quantum dots (QDs).

In the pumping process, active atoms experience acts of excitation at random moments in time \cite{Greenstein1995}. Generally, such act brings the atom to some state in the upper energy band. After the fast and nonradiative diffusion down this band, the atom encounters a long-lived laser transition where it interacts with the plasmonic mode of the nanolaser. Then, mixture of states $ \left| a \right\rangle  $ and $ \left| b \right\rangle  $ decays to the ground level, and the atom leaves the interaction. The atom is then ``worked out'', and is ready to reload.

The lower state $ \left| b \right\rangle  $ is usually chosen to be fast decaying to the ground state. In that case, direct transition from $ \left| a \right\rangle  $ to $ \left| g \right\rangle  $ can be neglected, and population of the level $ b $ is zero.

Lossless evolution of the plasmonic mode and the resonant atom transition, in the rotating-wave approximation, is governed by the Jaynes-Cummings Hamiltonian
\begin{equation}
H=\nu a^{+} a + \omega \sigma_z + g(\sigma_{-} a^{+} + \sigma_{+} a).
\label{eq21}
\end{equation}
Here, $ \nu $ and $ \omega $ are frequencies of the horseshoe and atom working transition. Lowering operator of the atom transition $ \sigma_- $ is connected with the inversion operator $ \sigma_z $ as follows: $ \sigma_z = \left[ \sigma_+,\sigma_- \right]  $. Coupling strength $ g $ depends on the position and orientation of the atom dipole moment $ \boldsymbol{\Pi} $ relative to the electric field \ref{eq16}:
\begin{equation}
g(r) = \sqrt{\dfrac{4 \pi \nu}{V}} \Pi_x  \sin(k_n z).
\label{eq22}
\end{equation}

Dynamics of the full density matrix (DM) is described by the equation
\begin{equation}
\dfrac{d \rho}{dt} = -i \left[ V, \rho \right] + \mathcal{L} \rho,
\label{eq23}
\end{equation}
where $ V = g(\sigma_{-} a^{+} + \sigma_{+} a) $ is Hamiltonian \ref{eq21} in the interaction picture, and $ \mathcal{L} $ is the Lindblad superoperator describing the dissipation in the atom and nanoresonator \cite{Milburn2008}:
\begin{equation}
\begin{split}
\mathcal{L} \rho = \sum_{i=1,2} \dfrac{\Gamma_i}{2} \left( 2\sigma_{-}^i \rho \sigma_{+}^i - \sigma_{+}^i \sigma_{-}^i \rho - \rho \sigma_{+}^i \sigma_{-}^i \right) + \\
\sum_{i=1,2} \gamma_i \left[ \sigma_{z}^i, \left[ \rho, \sigma_{z}^i \right] \right] + 
\dfrac{\kappa}{2} \left( 2a\rho a{+} - a^{+}a\rho - \rho a^{+}a \right) .
\end{split}
\label{eq24}
\end{equation}
Here, the first term corresponds to the atomic transitions, denoted by thin straight numbered arrows in the fig. \ref{p2}. The second term stands for the polarization decay of these transitions. The third term is responsible for the cavity losses. In the following, letters $ \Gamma $, $ \gamma $ and $ \sigma $ without index apply to the working atom transition.

Decay constants from \ref{eq24} together with the coupling strength $ g $ determine the time scales of the dynamics. As noted earlier, loss rates of the second transition $ \Gamma_2 $ and $ \gamma_2 $ are much larger than others and drop out. Then, only four scales $ \Gamma $, $ \gamma $, $ \kappa $ and $ g $ are important. $ \Gamma $ is the rate of the spontaneous emission into the free space and ranges from $ 10^{-8}~eV $ for dyes up to the $ 10^{-6}~eV $ for QDs. Dephasing $ \gamma $ is usually much faster and is about $ 10^{-2}~eV $ for QDs and $ 10^{-1}~eV $ for dyes. FOM of the plasmonic nanocavities ranges from $ 10 $ up to $ 10^3 $ and can be even higher at lower temperatures, or for the other types of cavities. That is, region of interest for $ \kappa $ is up to $ 10^{-1}~eV $. For the good-confined modes value of $ g $ of order $ 10^{-2}~eV $ seems to be the upper limit.

\subsection{\label{sec:solution}Solution}

In the present model, an excited atom interacts with the plasmonic mode individually, i.e. it does not know anything about the neighboring atoms. After this single interaction, atom can leave some of its initial energy in the nanolaser. If such events occur rarely, this enhancement is enough to sustain only, say, one quanta of energy in the resonator. More frequent ``kicks'', providing energy to compensate larger losses, may lead to the laser action. It is a logical extension of the shots concept, which we used in our earlier work \cite{Bogdanov2013} to describe the luminescence action in the nanolaser. On the language of the DM, the goal is to calculate the change of its elements as a result of a single excitation.

In the present model, we consider the case when the plasmon lifetime $ \kappa^{-1} $ is longer than time of interaction with the excited atom (we shall derive an explicit condition later on). In that case, calculation of the ``kick'' can be performed analytically. One should solve equation \ref{eq23} with an atom in initial pumped state and without the last term in \ref{eq24}. The plasmon dissipation enters only the coarse-grain master equation together with the constant flow of the kicks determined by the pumping rate. This approach is known as the maser approximation.

Now let us turn on to the details of the calculation. For the plasmon oscillator, we use the basis of the occupation numbers. Elements of the full DM are indexed as $ \rho_{\alpha n,\beta m} $, where $ \alpha,\beta \subset{a,b,g} $ indicate the state of the atom. Master equation \ref{eq23} then determines motion for these elements. In the maser approximation, it can be expressed in the form of a closed sets of equations:

\begin{widetext}
\begin{equation}
\label{eq25}
\begin{gathered}
 \dfrac{d}{dt} \rho_{an,am}  = -ig\left( \sqrt{n+1}\rho_{bn+1,am}-\sqrt{m+1}\rho_{an,bm+1}\right) - \Gamma_1 \rho_{an,am} \\
 \dfrac{d}{dt} \rho_{an,bm+1}  =-ig\left( \sqrt{n+1} \rho_{bn+1,bm+1} - \sqrt{m+1}\rho_{an,am} \right) -\gamma \rho_{an,bm+1} \\
 \dfrac{d}{dt} \rho_{bn+1,am}  =-ig\left( \sqrt{n+1} \rho_{an,am} - \sqrt{m+1}\rho_{bn+1,bm+1} \right) -\gamma \rho_{bn+1,am} \\
 \dfrac{d}{dt} \rho_{bn+1,bm+1}  =-ig\left( \sqrt{n+1} \rho_{an,bm+1} - \sqrt{m+1}\rho_{bn+1,am} \right) +\Gamma_1 \rho_{an+1,am+1} - \Gamma_2 \rho_{bn+1,bm+1} \\
 \dfrac{d}{dt} \rho_{gn+1,gm+1}  = \Gamma_2 \rho_{bn+1,bm+1}.
\end{gathered}
\end{equation}
\end{widetext}
Since $ \Gamma_2 $ is by far larger than other rates, equation for $ \rho_{bn+1,bm+1} $ can be excluded and one gets
\begin{widetext}
\begin{equation}
\label{eq26}
\begin{gathered}
 \dfrac{d}{dt} \rho_{an,am} = -ig\left( \sqrt{n+1}\rho_{bn+1,am}-\sqrt{m+1}\rho_{an,bm+1}\right) - \Gamma \rho_{an,am} \\
 \dfrac{d}{dt} \rho_{an,bm+1} = -ig \sqrt{m+1}\rho_{an,am} -\gamma \rho_{an,bm+1} \\
 \dfrac{d}{dt} \rho_{bn+1,am} = -ig \sqrt{n+1} \rho_{an,am} -\gamma \rho_{bn+1,am} \\
 \dfrac{d}{dt} \rho_{gn+1,gm+1} = -ig\left( \sqrt{n+1} \rho_{an,bm+1} - \sqrt{m+1}\rho_{bn+1,am} \right) + \Gamma \rho_{an+1,am+1}.
\end{gathered}
\end{equation}
\end{widetext}

Equations \ref{eq26} govern the internal dynamics of the kick. Initial value of the full DM is given by the outer product of the DM of the atom, prepared by the pumping mechanism, $ \rho_a^0 $, and that of plasmonic field oscillator at the moment of atom excitation, $ \rho_p^0 $: $ \rho^0=\rho_p^0\otimes \rho_a^0 $. In the simplest case, pumping brings the atom to the pure state $ \left| a \right\rangle $.  Initial conditions for the system \ref{eq26} then are $ \rho_{an,am}^0=\left( \rho_p^0 \right)_{n,m} $, $ \rho_{an,bm+1}^0=\rho_{bn+1,am}^0=\rho_{gn,gm}^0=0 $. In the course of time, evolution \ref{eq26} brings the atom to the ground state $ \left| g \right\rangle $, so that $ \rho_{an,am}\rightarrow 0 $ and $ \rho_{an,bm+1}\rightarrow 0,~\rho_{bn+1,am}\rightarrow 0 $. Final state of the plasmonic field can be obtained by tracing the ultimate DM over the atom variables:

\begin{equation}
\label{eq27}
\rho_p^{final} = \mathrm{Tr}_{atom} \left[ \rho^{final} \right] = \left\langle g\left| \rho^{final}\right|  g\right\rangle. 
\end{equation}

An expression for the kink of the DM then reads 

\begin{multline}
\label{eq28}
 \delta \rho_{n,m} = \left( \rho_p^{final} - \rho_p^0 \right)_{n,m} = \\
 \dfrac{2\sqrt{nm}}{n+m+2s} \rho_{n-1,m-1} - \dfrac{n+m+2}{n+m+2+2s} \rho_{n,m},
\end{multline}
where some indices are omitted for brevity: $ \rho_{n,m}=\left( \rho_p^0 \right)_{n,m} $, and we have introduced the constant $ s=\frac{\gamma \Gamma}{2g^2} $ which is the saturation number of quanta \cite{Kimble1998}. Process \ref{eq26} has a lifetime

\begin{equation}
\label{eq29}
\tau_{n,m} = \dfrac{2}{\Gamma+\gamma-\mathrm{Re} \sqrt{\left( \gamma - \Gamma \right)^2 - 4g^2 \left( n+m+2 \right) }},
\end{equation}
which is analog of the atom transit time in the theory of micromasers. The whole interaction process is characterized by the effective lifetime $ \tau_{eff} $, which should be less than the plasmon relaxation time. This is the application condition of our model ($ \Gamma \ll \gamma $):
\begin{equation}
\label{eq210}
\tau_{eff} \simeq \dfrac{2}{\gamma-\mathrm{Re} \sqrt{\gamma^2 - 8g^2 \left( \overline{n}+1 \right) }} < \kappa^{-1},
\end{equation}
where $ \overline{n} $ is the average number of quanta in the oscillator. 

An important quantity which can be traced from here is the fracture of the spontaneous emission going to the laser mode. This value equals to the coefficient before $ \rho_{n,m} $ in the right-hand side of \ref{eq28}, if $ n=m=0 $:
\begin{equation}
\label{eq211}
\beta = \dfrac{1}{1+s} = \dfrac{2g^2}{2g^2+\gamma\Gamma}.
\end{equation}
High value of $ \beta $ is distinctive feature of the microcavity lasers \cite{Bjork1994,Rice1994}. In this respect, plasmonic nanolasers are beyond comparison in the optical range. As follows from the previous section, realistic numbers give $ s \approx 10^{-4} $ and $ \beta $ as large as $ 0.9999 $.

Now we can construct the large-scale master equation. In line with the maser approach, we sum up the atom kinks, expr.\ref{eq28}, and add the plasmon dissipation term:
\begin{equation}
\label{eq212}
\dfrac{d}{dt} \rho = R \delta \rho + \left( \dfrac{d}{dt} \rho \right)_{loss}.
\end{equation}
Here, $ R $ is the rate at which new excited atoms enter the interaction with the mode, that is, pumping rate. For the latter summand, we use the standard expression \cite{Scully1997}:

\begin{equation}
\label{eq213}
\begin{aligned}
\left( \dfrac{d}{dt} \rho \right)&_{loss} 
 = - \overline{n}_{th} \dfrac{\kappa}{2} \left( aa^+ \rho - 2 a^+ \rho a + a^+ a \rho \right) \\
& - \left( \overline{n}_{th} + 1 \right) \dfrac{\kappa}{2} \left( a^+a \rho - 2 a \rho a^+ + aa^+ \rho \right),
\end{aligned}
\end{equation}
where $ \overline{n}_{th} $ is the number of quanta in thermal bath.

\begin{figure*}[htbp]
    \includegraphics[width=7in]{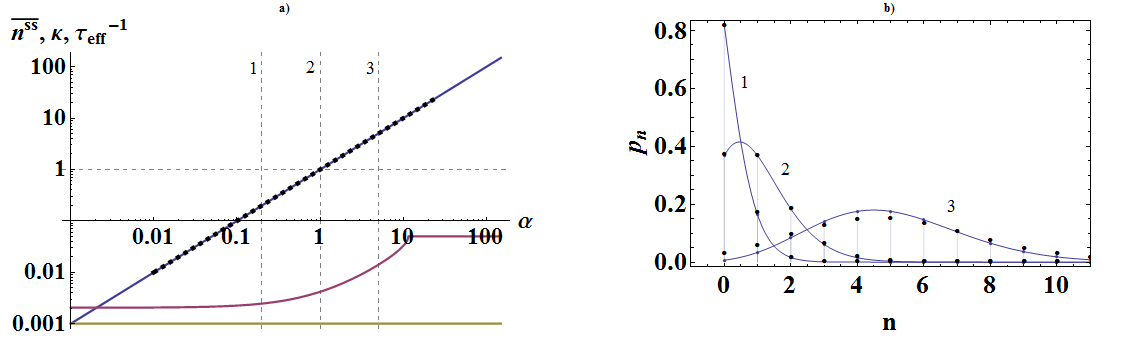}
    \caption{a) The average number of quanta \ref{eq33} (blue line - theory; black dashed - numeric results), and condition \ref{eq210}, with values in electronvolts: $ \kappa $ is yellow, $ \tau_{eff}^{-1} $ is purple.  b) statistics of quanta, at the pumpings marked by vertical dashed lines in the picture a (blue lines – theory; black dots – numeric calculation). $ s=0.5\times 10^{-4} $. Detailed paremeters: $ g=10^{-2}eV $, $ \Gamma = 10^{-7}eV $, $ \gamma=10^{-1}eV $, $ \kappa=10^{-3}eV $.}
    \label{p3}
\end{figure*}

\begin{figure*}[htbp]
    \includegraphics[width=7in]{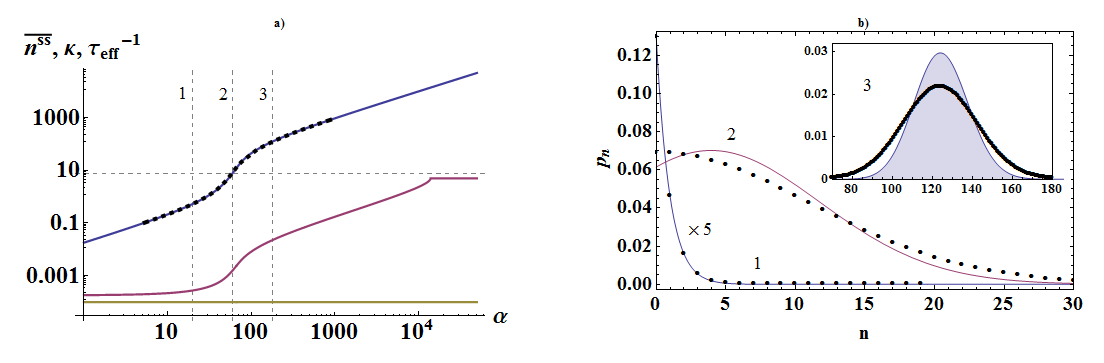}
    \caption{The same as fig.\ref{p3}, but for $ s=55.5 $ ($ g=3×10^{-5}eV $, $ \Gamma=10^{-5}eV $, $ \gamma=10^{-2}eV $, $ \kappa=10^{-7}eV $). Horizontal dashed line in the fig.a is the threshold number of quanta. Again, condition \ref{eq210} is fulfilled.}
    \label{p4}
\end{figure*}

\subsection{\label{sec:statistics}Plasmon statistics}

First, let us determine the steady state of the main diagonal of the DM. From eq.\ref{eq212}, we obtain a stationary condition for the diagonal elements $ p_n^{ss} = \rho_{n,n}^{steady~state} $, which are probabilities of finding $ n $ quanta in the mode (for simplicity, we take $ \overline{n}_{th} = 0 $):
\begin{equation}
\label{eq31}
R \left[ \dfrac{n}{n+s} p_{n-1}^{ss} - \dfrac{n+1}{n+1+s} p_{n}^{ss} \right] +\kappa \left[ \left( n+1 \right) p_{n+1}^{ss} - np_n^{ss} \right] = 0.
\end{equation}
Using the detailed balance condition, one finds the solution:
\begin{equation}
\label{eq32}
p_n^{ss} = A \dfrac{\alpha^n}{\Gamma\left[ n+s+1 \right]} \xrightarrow[s\rightarrow 0]{} \dfrac{e^{-n}\alpha^n}{n!}.
\end{equation}
Here, $ \Gamma $ is gamma function and $ A $ is a normalization factor. In case of strong coupling, when $ s\rightarrow 0 $, $ p_n^{ss} $ reduces to the Poisson distribution, regardless of the pumping rate. The total number of quanta in the mode is now straightforward:
\begin{equation}
\label{eq33}
n^{ss} = \sum_{n=0}^{\infty} np_n^{ss} = \alpha \beta \dfrac{{}_1F_1\left[ 2,2+s,\alpha\right] }{{}_1F_1\left[ 1,1+s,\alpha\right] } \xrightarrow[s\rightarrow 0]{} \alpha,
\end{equation}
where $ {}_1F_1 $ is the Kummer confluent hypergeometric function. Thus, the strong coupling between the laser mode and the active media leads to the linear dependence between $ n^{ss} $ and $ R $. Such behavior of laser is often referred as thresholdless, and is a commonplace in experiments on microlasers with high values of $ \beta $. 

Throughout the paper we shall illustrate our results using two sets of parameters. The first one stands for the real plasmonic nanolaser with Q-factor of $ 10^3 $ and strong coupling: $ g=10^{-2}eV $, which is the limit of $ s\rightarrow 0 $ and $ \beta\rightarrow 1 $. The second one is closer to the nanolaser based on the photonic crystal or microsphere cavity. Our model covers these cases without any changes since the laser mode quantization for them leads to the same results of the section \ref{sec:plq}. Such cavities usually have much higher finesses and larger mode volumes, which results in the weaker coupling.  We take $ g=3\times 10^{-5}eV $  and $ Q=10^7 $, which gives $ s=55.5 $ and $ \beta=0.017 $.

Figure \ref{p3} illustrates the results \ref{eq32} and \ref{eq33} in the first case. In the fig.\ref{p3}a, the blue line shows the dependence between the average number of quanta and the pumping parameter $ \alpha $. It agrees perfectly with the result of the numeric simulation (black dashed line), which does not use the maser approximation. It is linear, as long as all the excitation energy goes to the mode, either in the spontaneous or stimulated way. Still, care should be taken to satisfy the restriction \ref{eq210}. Yellow line indicates the present value of $ \kappa $ and purple line stands for $ \tau_{eff}^{-1} $. Since the purple line is above yellow one everywhere, condition \ref{eq210} is met and model is adequate for the parameters used.

Blue lines in the picture \ref{p3}b show the statistics of plasmons \ref{eq32} for three different pumpings, marked by dashed lines in \ref{p3}a. Black dots stand for the distributions obtained in the numeric calculation, which generally follow the theoretical prediction. We shall explore the differences in more detail in discussion of the second-order coherence. 

Distributions are close to Poissonian in a good approximation, as expected in the limit $ \beta \rightarrow 1 $: the first set of parameters gives $ \beta \approx 0.99995 $. Note, that at the second cross-section the number of quanta equals unity, and $ p_0^{ss}=p_1^{ss} $. These conitions express two different definitions of the laser threshold (\cite{Bjork1994,Bykov1994,Scully1997}), which coincide in the limit in question.

Figure \ref{p4} shows the same things for the second set of parameters. Theoretical prediction of the number of quanta is surprisingly good, considering serious discrepancies in the statistics. Note, that the average number of quanta in the fig.\ref{p4}a has a familiar ``kink'' at the threshold region, which indicates a relatively low value of $ \beta $, which amounts now to $ \approx 0.017 $. Limits in \ref{eq32} and \ref{eq33} are no more relevant. As a result, distribution of quanta is different from the Poissonian, it is wider. Well above threshold, though, statistics is approaching that of the coherent state, as appears from the inset in fig.\ref{p4}b. Section, marked by the second dashed line in \ref{p4}a is drawn in the point of maximum slope of the blue line. At this point, number of quanta equals $ \sqrt{s} $ (horizontal line in \ref{p4}a, as expected for classical good-cavity lasers \cite{Carmichael1999}.
 
It is easily seen, that all the DM elements, except those on the main diagonal, are zero in the steady state of \ref{eq212}. This state of the field is mixture of the pure states with the uniform distribution of phases. When elements of the main diagonal follow the Poissonian distribution, this mixture is often referred as the coherent state, though it is obviously not the case. Actually, DM of that kind just represents our state of knowledge of the field state with the total ignorance of its phase. Of course, it does not affect any of its coherence properties \cite{Glauber2007}. Interestingly, that the state of affairs changes, if atoms are pumped into the pure superposition of the working states. Then, phase of the field becomes partly predictable, which leads to the nonzero high-order diagonals in the DM \cite{Bykov1994}.

\section{\label{Coherence}Coherence properties of the plasmonic radiation}

What we are interested above all, are, of course, properties of the outgoing radiation. In our model, it is represented by the continuum of electromagnetic modes in the free space, constituting the reservoirs to which active atoms and the laser mode are coupled. Interaction with these modes is traced out and enters our model via the term \ref{eq213}. There, summands including the radiative decay are those proportional to $ \Gamma_1 $ and $ \kappa $. These constant actually account for the nonradiative decay also (ohmic losses in metal, for example). The former, which is spontaneous decay of the working atom transition into the free space, is almost entirely suppressed when $ \beta\approx 1 $. Therefore, we shall concentrate on the radiative part of the nanoresonator decay. 

It can be shown, that the corresponding DM of the electromagnetic oscillator in the far-field zone is exactly the same as for the radiating mode, up to the corresponding time delay and overall reduction of the intensity \cite{Scully1997}. Accordingly, positive and negative frequency parts of the electric far-field are proportional to the plasmonic operators $ a $ and $ a^+ $. Methods, developed in the preceding sections, allow us to obtain the DM of interest; results on the quantum statistic thus hold for the outgoing radiation. That is, we have all instruments to explore the radiation of the plasmonic mode.

The primary question to this section is: whether the radiation of the nanolaser is coherent? It is strongly connected with the definition of the laser threshold which became ambiguous as $ \beta $ is approaching unity. In this section, we shall address this issue from the framework of our theory. We concentrate first on the analysis of the first-order coherence, which will allow us to determine the spectrum and the linewidth of the plasmon radiation, and then turn to the study of the second-order effects.

\subsection{First-order coherence}

One of the widespread ideas about coherence is identifying it with the monochromaticity of radiation. This is a misleading view, although containing a grain of truth. In fact, monochromaticity does lead to the coherence of radiation, but only of the first order \cite{Glauber2007} and only for the stationary sources. Stationarity is obvious for the measurements in the CW regime but is questionable in the experiments dealing with the pico- and femto- second pulses. Here, we shall study the first-order coherence of the nanolaser radiation in the CW case, when the pumping rate can be thought as a constant.

Both of these ideas rely on the concept of the first-order correlation function. Its normalized version, called the first-order degree of coherence, is defined as \cite{Glauber2007,Scully1997}
\begin{equation}
\label{eq41}
g^1 (\tau) = \dfrac{\left< a^+(t) a(t+\tau) \right>}{\left< a^+ a \right>}.
\end{equation}
Here, operators $ a $ and $ a^+ $ mean the same as before, but now in Heisenberg representation are functions of time, taken at the same point in space. Angle brackets denote the quantum statistical average. Function \ref{eq41} is valued between zero and unity. They correspond to the minimal and the maximal visibility of the interference fringes in the Young's experiment, which is the simplest procedure of observation the first-order coherence effect. Spectrum of the stationary field is the Fourier transform of this function \cite{Scully1997}:
\begin{equation}
\label{eq42}
S(\omega)=\mathrm{Re}\int_{0}^{\infty} d\tau g^1 (\tau) e^{i \omega \tau}.
\end{equation}

That is, the challenge is to calculate the expression \ref{eq42} with the two-time average in the numerator. As long as the motion of the DM, \ref{eq212}, is the Markov process, \ref{eq41} can be rewritten as follows \cite{Gardiner2004}:
\begin{equation}
\label{eq43}
g^1(\tau) = \dfrac{\mathrm{Tr} \left[ aU(t+\tau,t) \left(\rho(t) a^+\right) \right]}{\left< a^+ a \right>},
\end{equation}
where $ U $ is operator of the evolution \ref{eq212}, and $ \rho^{ss}(t) $ is steady state of the DM. Multiplication by $ a^+ $ shifts the values of zero diagonal to the first one. Function $ g^1(\tau) $ is thus determined by evolution of the first diagonal of the DM, which follows this same law \ref{eq212} and leads to the zero steady state, as noted earlier. To obtain the decay dynamics it is convenient to write its equation of motion in the following form:

\begin{widetext}
\begin{equation}
\label{eq44}
\begin{gathered}
\frac{d}{dt}\rho_{n,n+1} = R \left[ \frac{\sqrt{n(n+1)}}{n+1/2+s}\rho_{n,n-1} - \frac{(n+1)(n+2)}{n+3/2+s}\rho_{n,n+1} \right] + \kappa \left[\sqrt{(n+1)(n+2)}\rho_{n+1,n+2} - \sqrt{n(n+1)}\rho_{n,n+1}\right] - \mu_n \rho_{n,n+1},
\end{gathered}
\end{equation}
where
\begin{equation}
\label{eq45}
\begin{gathered}
\mu_n = \kappa \left(n+\frac{1}{2}-\sqrt{n(n+1)}\right) + R \dfrac{n+3/2-\sqrt{(n+1)(n+2)}}{n+3/2+s}.
\end{gathered}
\end{equation}
\end{widetext}
Eq. \ref{eq44} without the last term has a steady state with
\begin{equation}
\label{eq46}
\rho_{n,n+1}^{ss}\propto \dfrac{\alpha^n}{\Gamma\left[n+3/2+s\right]}.
\end{equation}
The last term in \ref{eq44} is a small disturbance relative to the rest of the system, and expression \ref{eq45} is thus the decay rate of $ \rho_{n,n+1} $, provided that statistics of these elements follows the quasi-stationary distribution \ref{eq46}. 

Initial point for the evolution \ref{eq43} is $ \rho^{ss}a^+ $, which has the nonzero elements only on the diagonal next to the main, whose distribution is
\begin{equation}
\label{eq47}
\rho_{n,n+1}^* = \sqrt{n+1} p_{n+1}^{ss} \propto \dfrac{\sqrt{n+1}\alpha^n}{\Gamma\left[n+3/2+s\right]}.
\end{equation}
Form of this function is close to that of \ref{eq46}, and in a good approximation can be regarded as the same. That is, initial state for evolution $ U $ in \ref{eq43} is already the steady one. Actually, it holds the same form \ref{eq46} throughout the decay process since the disturbance in \ref{eq44} is small.

If $ \overline{n} \gg 1 $, that is, in the limit of strong pumping, function \ref{eq46} is peaked around this large value, the only region where a substantial decay occurs; dependence of $ n $ in \ref{eq45} then can be neglected and one obtains
\begin{equation}
\label{eq48}
\begin{aligned}
& g^1(\tau) =  \exp \left( -\mu_{\overline{n}} \tau\right) ,\\
& \mu_{\overline{n}} = \frac{\kappa}{8\overline{n}} + \frac{R}{8\overline{n}^2}+o\left( \frac{1}{\overline{n}^2}\right) \xrightarrow[\overline{n}\rightarrow \infty]{} \frac{\kappa}{4\alpha}.
\end{aligned}
\end{equation}
As follows from \ref{eq42}, spectrum then has a Lorentzian shape with the traditional linewidth \ref{eq48}. Far more interesting is the spectrum in case of $ n\lesssim 1 $. In this domain, the approximation used in for the eq \ref{eq48} goes wrong, as does the common model of the phase diffusion. Next, we shall dig a bit deeper and obtain an expression for $ g^1 $ which in valid for arbitrary values of $ \overline{n} $.

In the evolution $ U $ (eq.\ref{eq43}), the decay with ``statistics'' \ref{eq45} tends to distort the quasi-stationary distribution. This action is compensated at the price of small deviation of the actual distribution of elements $ \rho_{n,n+1} $ from the \ref{eq46}. Therefore, the decay of the trace in \ref{eq43} is indeed exponential for the $ \rho^{ss} $ with any $ \overline{n} $. Its rate is then given by averaging of \ref{eq45} over the steady state distribution \ref{eq46}. One obtains
\begin{equation}
\label{eq49}
\begin{aligned}
& g^1(\tau) = \exp(-\mu_{eff} \tau),\\
& \mu_{eff}=\frac{1}{\overline{n}} \sum_{n=0}^{\infty} (n+1)p_{n+1}^{ss}\mu_n.
\end{aligned}
\end{equation}
That is, we expect the spectrum to have the Lorentzian shape with HWHM equal to $ \mu_{eff} $.

\begin{figure*}[htbp]
    \includegraphics[width=7in]{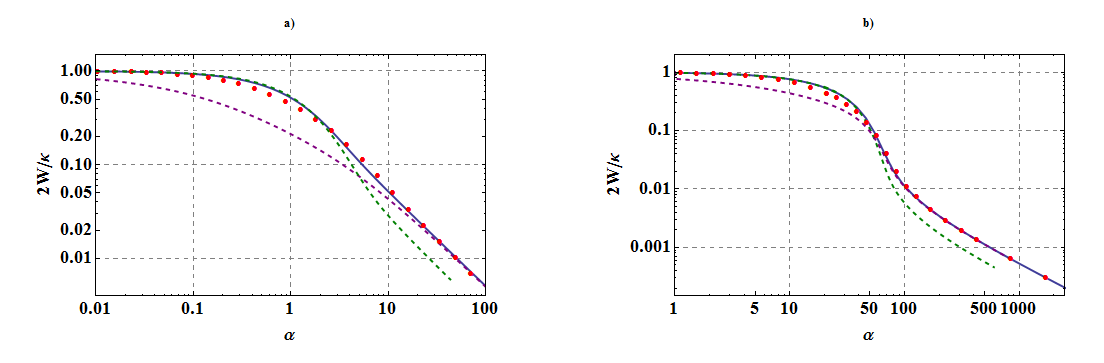}
    \caption{HWHM of the laser mode radiation, normalized to that of the free resonator. Blue lines: $ \mu_{eff} $. Purple lines: $ \mu_{\overline{n}} $. Green lines: expr. \ref{eq410}. Red dots: numeric simulation. Plots a and b stand for the sets of parameters corresponding to plasmonic- and micro- lasers}
    \label{p5}
\end{figure*}

Sum in \ref{eq49} cannot be taken exactly in general. In the case $ \alpha\lesssim 1 $, opposite to that of \ref{eq48}, analytical approximation can also be obtained. Here, $ \mu_n $ (\ref{eq45}) can be fitted by the function $ \frac{\kappa}{8n} $, starting from $ n=1 $, and summation in \ref{eq49} then can be performed. Resulting expression is rather cumbersome, and here we show only its limit of small $ s $:
\begin{equation}
\label{eq410}
\begin{aligned}
& \mu_{eff} \xrightarrow[s\rightarrow 0]{} \\
& \frac{\kappa}{2} e^{-\alpha} \left[1+\alpha\left(2-\frac{4}{3}\sqrt{2}\right)+\frac{\mathrm{Ei}(\alpha)-\ln\alpha+\gamma_e}{4}\right].
\end{aligned}
\end{equation}
Here, $ \mathrm{Ei} $ is exponential integral funciton and $ \gamma_e\approx 0.577 $ is the Euler's constant. Figure \ref{p5} shows these widths plotted versus $ \alpha $. Picture $ a $ stands for the case of small-$ s $ laser, with parameters as for fig.\ref{p3}. $ \mu_{eff} $ from \ref{eq49} is plotted as a blue line. For comparison, we plotted the corresponding results obtained in the numeric solution of the master equation without the maser approximation (red dots).

Essentially, in the fig.\ref{p5}a one sees the linewidth proportional to the inverse number of quanta (recall that $ n^{ss}\propto \alpha $ ), but limited from above by the cold-cavity linewidth, $ \kappa/2 $. These two regimes, corresponding to the predominance of the spontaneous or stimulated energy transfer between active medium and the resonator, meet when $ n^{ss}\approx 1 $. Expression \ref{eq49} agrees with the numeric results perfectly, except the transition region, where the discrepancy reaches 20 percent. The purple line is the value of $ \mu_{\overline{n}} $ , which approximates $ \mu_{eff} $ in the limit of strong pumping. The opposite case is covered by the approximation \ref{eq410}, the green dashed line.

Figure \ref{p5}b is the same for the microlaser-like system, with parameters from the fig.\ref{p4}. We observe the kink-like element, resembling that in the plasmon number line in fig.\ref{p4}a. The purple line, or the ``classical'' linewidth, is perfect approximation after the kink, or classically, above the threshold. The slow-pumping approximation (green line) works even better than expected and holds up to $ \alpha\approx s $. Actually, we cover the linewidth analytically almost in the whole range of $ \alpha $.

\begin{figure}[htbp]
    \includegraphics[width=3.2in]{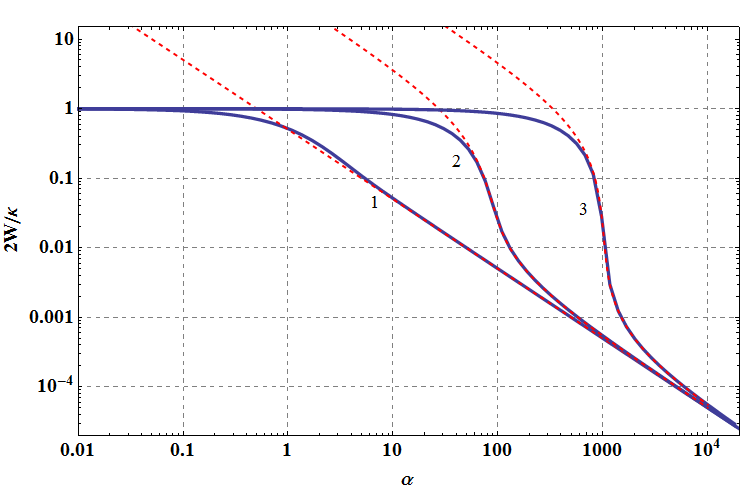}
    \caption{Linewidths \ref{eq49} plotted for systems with $ s=10^{-2},80,10^3 $ (blue lines). Marked by numbers 1-3, respectively. Red dashed lines are proportional to the inverse numbers of quanta.}
    \label{p6}
\end{figure}

To illustrate the logic of the spectra behavior clearer, we plot it versus $ \alpha $ for several values of $ s $ on the same graph (figure \ref{p6}).

Red dashes, which essentially are the inverse numbers of quanta, lay precisely on the linewidth lines at large $ \alpha $. The departure starts when the linewidths begin to feel their ceiling, the cold-cavity value. This happens right after the kink-like intensity reduction (if any) occurs, when $ \alpha \approx s $ (for large $ s $). Origin of the lines in the fig.\ref{p5} now became transparent: graph \ref{p5}a has the form of the first line in fig.\ref{p6}, while \ref{p5}b is closer to the second.

Picture \ref{p6} also shows how the classical concept of the laser threshold, based on the line narrowing, fades away as we leave the thermodynamic limit. Line 3 demonstrates a two orders of magnitude leap of the linewidth when the pumping rate doubles. No such thing is expected to happen in the plasmonic, thresholdless nanolaser. In that case, the only characteristic point which indicates the switch of the working regimes is $ \alpha=n^{ss}=1 $. At that point, the linewidths roughly halves its initial value. The exact number is $ \approx 0.53 $.

\subsection{Second-order coherence}

Now we turn to the analysis of the second-order coherence of the plasmon radiation. It is based on the features of the second-order correlation function\cite{Glauber2007,Scully1997}
\begin{equation}
\label{eq411}
g^2(\tau) = \dfrac{\left< a^+ (t)a^+ (t+\tau)a(t+\tau)a(t) \right>}{\left< a^+ a \right>^2}.
\end{equation}

\begin{figure*}[htbp]
    \includegraphics[width=7in]{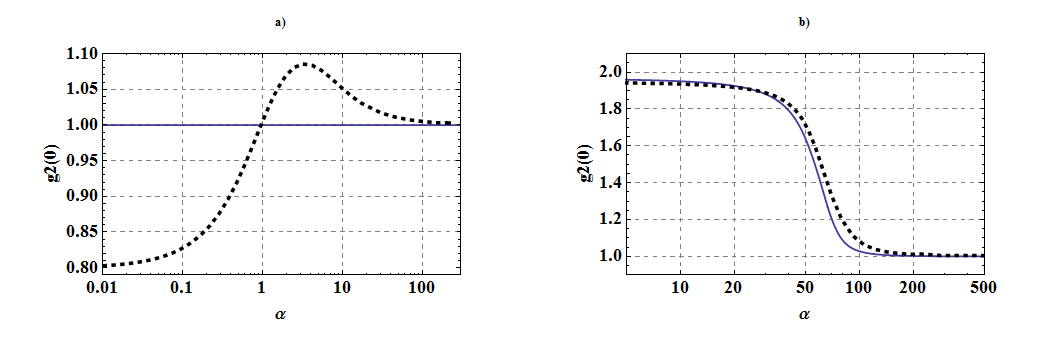}
    \caption{Theoretical (blue lines) and numeric (black dashed) values of $ g^2 (0) $ versus $ \alpha $, for the first (a) and second (b) sets of parameters.}
    \label{p7}
\end{figure*}

Of greatest importance is its value for $ \tau=0 $, which allows one to determine the nature of radiation considered. Signature of the macroscopic laser radiation well above threshold is $ g^2 (0)=1 $, which characterizes the field in the mixture of coherent states, discussed in the section \ref{sec:statistics}. As follows from eq.\ref{eq411}, $ g^2 (0) $ is an intrinsic feature of the statistics of quanta, and can be readily calculated, given the distribution \ref{eq32}. Calculation results are shown in the figure \ref{p7}. 

As appears from the graph a, our model misses essential features of the statistics of quanta. Numeric result shows that the statistics is not Poissonian everywhere, as follows from the theory. Instead, it is super-Poissonian if $ \alpha>1 $ and sub-Poissonian if $ \alpha<1 $. Discrepancy manifests itself in the fig.\ref{p3}b, where the third dashed line stands nearly for the peak of the numeric line in \ref{p7}a. Interestingly, that at the point $ \alpha=1 $, numeric line returns to the predicted value of unity. The reason why the theory fails to predict $ g^2 $ is that the condition \ref{eq210} is not strong enough. As we have seen, it works well, if we are interested in the spectrum of the radiation or the number of quanta. But inaccurate neglecting of the plasmon dissipation during the atom interaction time happens to be ruinous for the fine features of the statistics. To preserve them, sign $ < $ in \ref{eq210} should really be replaced by $ \ll $.

For the second set of parameters, condition \ref{eq210} is satisfied better, and consequently the figure \ref{p7}b demonstrates a much better agreement. At weak pumpings, the theoretical line starts from the value
\begin{equation}
\label{eq412}
g^2 (0,\alpha\rightarrow 0)=2\dfrac{s+1}{s+2}.
\end{equation}
For large $ s $, it reduces to the value of two, which indicates the thermal state of light, in agreement with the results of the traditional laser theory. Kink in the $ g^2 (0) $ line occurs near the same value $ \alpha\approx s $, as it does in figures \ref{p4},\ref{p5},\ref{p6}. At higher $ \alpha $, the value of  $ g^2 (0) $ tends to unity, indicating the transition to the coherent state of light.

\section{Conclusion}
We have developed the density matrix model of the nanolaser, based on the maser approximation of the laser operation. We have shown that the plasmonic nanolaser is a truly thresholdless device, due to the extremely strong coupling of the plasmonic mode and the active medium, which result in values of $ \beta $, possibly as large as $ 0.9999 $. We obtain the analytical expressions for the number of quanta and the linewidth of the laser radiation, which are valid at any pumping rate and agree with the numeric simulations. Our theory also covers a broad range of the high-Q microcavity lasers. 

Plasmonic nanolasers, with respect to the Q-factor, are close to the edge of applicability of the maser model. The latter allows one to analyze the intensity of laser radiation and first-order coherence effects, but misses the fine features of the plasmon statistics, which determine degree of the second-order coherence. We conclude, that for the further theoretical DM-based investigation of the bad-cavity plasmonic nanolasers, one should go beyond the maser approximation.

\bibliography{library}

\begin{thebibliography}{36}%
\makeatletter
\providecommand \@ifxundefined [1]{%
 \@ifx{#1\undefined}
}%
\providecommand \@ifnum [1]{%
 \ifnum #1\expandafter \@firstoftwo
 \else \expandafter \@secondoftwo
 \fi
}%
\providecommand \@ifx [1]{%
 \ifx #1\expandafter \@firstoftwo
 \else \expandafter \@secondoftwo
 \fi
}%
\providecommand \natexlab [1]{#1}%
\providecommand \enquote  [1]{``#1''}%
\providecommand \bibnamefont  [1]{#1}%
\providecommand \bibfnamefont [1]{#1}%
\providecommand \citenamefont [1]{#1}%
\providecommand \href@noop [0]{\@secondoftwo}%
\providecommand \href [0]{\begingroup \@sanitize@url \@href}%
\providecommand \@href[1]{\@@startlink{#1}\@@href}%
\providecommand \@@href[1]{\endgroup#1\@@endlink}%
\providecommand \@sanitize@url [0]{\catcode `\\12\catcode `\$12\catcode
  `\&12\catcode `\#12\catcode `\^12\catcode `\_12\catcode `\%12\relax}%
\providecommand \@@startlink[1]{}%
\providecommand \@@endlink[0]{}%
\providecommand \url  [0]{\begingroup\@sanitize@url \@url }%
\providecommand \@url [1]{\endgroup\@href {#1}{\urlprefix }}%
\providecommand \urlprefix  [0]{URL }%
\providecommand \Eprint [0]{\href }%
\providecommand \doibase [0]{http://dx.doi.org/}%
\providecommand \selectlanguage [0]{\@gobble}%
\providecommand \bibinfo  [0]{\@secondoftwo}%
\providecommand \bibfield  [0]{\@secondoftwo}%
\providecommand \translation [1]{[#1]}%
\providecommand \BibitemOpen [0]{}%
\providecommand \bibitemStop [0]{}%
\providecommand \bibitemNoStop [0]{.\EOS\space}%
\providecommand \EOS [0]{\spacefactor3000\relax}%
\providecommand \BibitemShut  [1]{\csname bibitem#1\endcsname}%
\let\auto@bib@innerbib\@empty
\bibitem [{\citenamefont {Bergman}\ and\ \citenamefont
  {Stockman}(2003)}]{Bergman2003}%
  \BibitemOpen
  \bibfield  {author} {\bibinfo {author} {\bibfnamefont {D.}~\bibnamefont
  {Bergman}}\ and\ \bibinfo {author} {\bibfnamefont {M.}~\bibnamefont
  {Stockman}},\ }\href {\doibase 10.1103/PhysRevLett.90.027402} {\bibfield
  {journal} {\bibinfo  {journal} {Physical Review Letters}\ }\textbf {\bibinfo
  {volume} {90}},\ \bibinfo {pages} {1} (\bibinfo {year} {2003})}\BibitemShut
  {NoStop}%
\bibitem [{\citenamefont {Noginov}\ \emph
  {et~al.}(2006{\natexlab{a}})\citenamefont {Noginov}, \citenamefont {Zhu},
  \citenamefont {Bahoura}, \citenamefont {Adegoke}, \citenamefont {Small},
  \citenamefont {Ritzo}, \citenamefont {Drachev},\ and\ \citenamefont
  {Shalaev}}]{Noginov2006}%
  \BibitemOpen
  \bibfield  {author} {\bibinfo {author} {\bibfnamefont {M.}~\bibnamefont
  {Noginov}}, \bibinfo {author} {\bibfnamefont {G.}~\bibnamefont {Zhu}},
  \bibinfo {author} {\bibfnamefont {M.}~\bibnamefont {Bahoura}}, \bibinfo
  {author} {\bibfnamefont {J.}~\bibnamefont {Adegoke}}, \bibinfo {author}
  {\bibfnamefont {C.}~\bibnamefont {Small}}, \bibinfo {author} {\bibfnamefont
  {B.}~\bibnamefont {Ritzo}}, \bibinfo {author} {\bibfnamefont
  {V.}~\bibnamefont {Drachev}}, \ and\ \bibinfo {author} {\bibfnamefont
  {V.}~\bibnamefont {Shalaev}},\ }\href {\doibase 10.1007/s00340-006-2401-0}
  {\bibfield  {journal} {\bibinfo  {journal} {Applied Physics B}\ }\textbf
  {\bibinfo {volume} {86}},\ \bibinfo {pages} {455} (\bibinfo {year}
  {2006}{\natexlab{a}})}\BibitemShut {NoStop}%
\bibitem [{\citenamefont {Noginov}\ \emph
  {et~al.}(2006{\natexlab{b}})\citenamefont {Noginov}, \citenamefont {Zhu},
  \citenamefont {Bahoura}, \citenamefont {Adegoke}, \citenamefont {Small},
  \citenamefont {Ritzo}, \citenamefont {Drachev},\ and\ \citenamefont
  {Shalaev}}]{Noginov2006a}%
  \BibitemOpen
  \bibfield  {author} {\bibinfo {author} {\bibfnamefont {M.~a.}\ \bibnamefont
  {Noginov}}, \bibinfo {author} {\bibfnamefont {G.}~\bibnamefont {Zhu}},
  \bibinfo {author} {\bibfnamefont {M.}~\bibnamefont {Bahoura}}, \bibinfo
  {author} {\bibfnamefont {J.}~\bibnamefont {Adegoke}}, \bibinfo {author}
  {\bibfnamefont {C.~E.}\ \bibnamefont {Small}}, \bibinfo {author}
  {\bibfnamefont {B.~a.}\ \bibnamefont {Ritzo}}, \bibinfo {author}
  {\bibfnamefont {V.~P.}\ \bibnamefont {Drachev}}, \ and\ \bibinfo {author}
  {\bibfnamefont {V.~M.}\ \bibnamefont {Shalaev}},\ }\href
  {http://www.ncbi.nlm.nih.gov/pubmed/17001387} {\bibfield  {journal} {\bibinfo
   {journal} {Optics letters}\ }\textbf {\bibinfo {volume} {31}},\ \bibinfo
  {pages} {3022} (\bibinfo {year} {2006}{\natexlab{b}})}\BibitemShut {NoStop}%
\bibitem [{\citenamefont {Chang}\ \emph {et~al.}(2007)\citenamefont {Chang},
  \citenamefont {S\o~rensen}, \citenamefont {Hemmer},\ and\ \citenamefont
  {Lukin}}]{Chang2007}%
  \BibitemOpen
  \bibfield  {author} {\bibinfo {author} {\bibfnamefont {D.}~\bibnamefont
  {Chang}}, \bibinfo {author} {\bibfnamefont {A.}~\bibnamefont {S\o~rensen}},
  \bibinfo {author} {\bibfnamefont {P.}~\bibnamefont {Hemmer}}, \ and\ \bibinfo
  {author} {\bibfnamefont {M.}~\bibnamefont {Lukin}},\ }\href {\doibase
  10.1103/PhysRevB.76.035420} {\bibfield  {journal} {\bibinfo  {journal}
  {Physical Review B}\ ,\ \bibinfo {pages} {1}} (\bibinfo {year}
  {2007})}\BibitemShut {NoStop}%
\bibitem [{\citenamefont {Chang}\ \emph {et~al.}(2006)\citenamefont {Chang},
  \citenamefont {S\o~rensen}, \citenamefont {Hemmer},\ and\ \citenamefont
  {Lukin}}]{Chang2006}%
  \BibitemOpen
  \bibfield  {author} {\bibinfo {author} {\bibfnamefont {D.}~\bibnamefont
  {Chang}}, \bibinfo {author} {\bibfnamefont {A.}~\bibnamefont {S\o~rensen}},
  \bibinfo {author} {\bibfnamefont {P.}~\bibnamefont {Hemmer}}, \ and\ \bibinfo
  {author} {\bibfnamefont {M.}~\bibnamefont {Lukin}},\ }\href {\doibase
  10.1103/PhysRevLett.97.053002} {\bibfield  {journal} {\bibinfo  {journal}
  {Physical review letters}\ }\textbf {\bibinfo {volume} {053002}},\ \bibinfo
  {pages} {1} (\bibinfo {year} {2006})}\BibitemShut {NoStop}%
\bibitem [{\citenamefont {Kimble}(1998)}]{Kimble1998}%
  \BibitemOpen
  \bibfield  {author} {\bibinfo {author} {\bibfnamefont {H.~J.}\ \bibnamefont
  {Kimble}},\ }\href {\doibase 10.1238/Physica.Topical.076a00127} {\bibfield
  {journal} {\bibinfo  {journal} {Physica Scripta}\ }\textbf {\bibinfo {volume}
  {T76}},\ \bibinfo {pages} {127} (\bibinfo {year} {1998})}\BibitemShut
  {NoStop}%
\bibitem [{\citenamefont {Akimov}\ \emph {et~al.}(2007)\citenamefont {Akimov},
  \citenamefont {Mukherjee}, \citenamefont {Yu}, \citenamefont {Chang},
  \citenamefont {Zibrov}, \citenamefont {Hemmer}, \citenamefont {Park},\ and\
  \citenamefont {Lukin}}]{Akimov2007}%
  \BibitemOpen
  \bibfield  {author} {\bibinfo {author} {\bibfnamefont {a.~V.}\ \bibnamefont
  {Akimov}}, \bibinfo {author} {\bibfnamefont {A.}~\bibnamefont {Mukherjee}},
  \bibinfo {author} {\bibfnamefont {C.~L.}\ \bibnamefont {Yu}}, \bibinfo
  {author} {\bibfnamefont {D.~E.}\ \bibnamefont {Chang}}, \bibinfo {author}
  {\bibfnamefont {a.~S.}\ \bibnamefont {Zibrov}}, \bibinfo {author}
  {\bibfnamefont {P.~R.}\ \bibnamefont {Hemmer}}, \bibinfo {author}
  {\bibfnamefont {H.}~\bibnamefont {Park}}, \ and\ \bibinfo {author}
  {\bibfnamefont {M.~D.}\ \bibnamefont {Lukin}},\ }\href {\doibase
  10.1038/nature06230} {\bibfield  {journal} {\bibinfo  {journal} {Nature}\
  }\textbf {\bibinfo {volume} {450}},\ \bibinfo {pages} {402} (\bibinfo {year}
  {2007})}\BibitemShut {NoStop}%
\bibitem [{\citenamefont {Dominguez}\ \emph {et~al.}(2013)\citenamefont
  {Dominguez}, \citenamefont {Regan}, \citenamefont {Bernussi},\ and\
  \citenamefont {{Grave de Peralta}}}]{Dominguez2013}%
  \BibitemOpen
  \bibfield  {author} {\bibinfo {author} {\bibfnamefont {D.}~\bibnamefont
  {Dominguez}}, \bibinfo {author} {\bibfnamefont {C.~J.}\ \bibnamefont
  {Regan}}, \bibinfo {author} {\bibfnamefont {a.~a.}\ \bibnamefont {Bernussi}},
  \ and\ \bibinfo {author} {\bibfnamefont {L.}~\bibnamefont {{Grave de
  Peralta}}},\ }\href {\doibase 10.1063/1.4792305} {\bibfield  {journal}
  {\bibinfo  {journal} {Journal of Applied Physics}\ }\textbf {\bibinfo
  {volume} {113}},\ \bibinfo {pages} {073102} (\bibinfo {year}
  {2013})}\BibitemShut {NoStop}%
\bibitem [{\citenamefont {Noginov}\ \emph {et~al.}(2009)\citenamefont
  {Noginov}, \citenamefont {Zhu}, \citenamefont {Belgrave}, \citenamefont
  {Bakker}, \citenamefont {Shalaev}, \citenamefont {Narimanov}, \citenamefont
  {Stout}, \citenamefont {Herz}, \citenamefont {Suteewong},\ and\ \citenamefont
  {Wiesner}}]{Noginov2009}%
  \BibitemOpen
  \bibfield  {author} {\bibinfo {author} {\bibfnamefont {M.}~\bibnamefont
  {Noginov}}, \bibinfo {author} {\bibfnamefont {G.}~\bibnamefont {Zhu}},
  \bibinfo {author} {\bibfnamefont {A.~M.}\ \bibnamefont {Belgrave}}, \bibinfo
  {author} {\bibfnamefont {R.}~\bibnamefont {Bakker}}, \bibinfo {author}
  {\bibfnamefont {V.~M.}\ \bibnamefont {Shalaev}}, \bibinfo {author}
  {\bibfnamefont {E.~E.}\ \bibnamefont {Narimanov}}, \bibinfo {author}
  {\bibfnamefont {S.}~\bibnamefont {Stout}}, \bibinfo {author} {\bibfnamefont
  {E.}~\bibnamefont {Herz}}, \bibinfo {author} {\bibfnamefont {T.}~\bibnamefont
  {Suteewong}}, \ and\ \bibinfo {author} {\bibfnamefont {U.}~\bibnamefont
  {Wiesner}},\ }\href {\doibase 10.1038/nature08318} {\bibfield  {journal}
  {\bibinfo  {journal} {Nature}\ }\textbf {\bibinfo {volume} {460}},\ \bibinfo
  {pages} {1110} (\bibinfo {year} {2009})}\BibitemShut {NoStop}%
\bibitem [{\citenamefont {Oulton}\ \emph {et~al.}(2009)\citenamefont {Oulton},
  \citenamefont {Sorger}, \citenamefont {Zentgraf}, \citenamefont {Ma},
  \citenamefont {Gladden}, \citenamefont {Dai}, \citenamefont {Bartal},\ and\
  \citenamefont {Zhang}}]{Oulton2009}%
  \BibitemOpen
  \bibfield  {author} {\bibinfo {author} {\bibfnamefont {R.~F.}\ \bibnamefont
  {Oulton}}, \bibinfo {author} {\bibfnamefont {V.~J.}\ \bibnamefont {Sorger}},
  \bibinfo {author} {\bibfnamefont {T.}~\bibnamefont {Zentgraf}}, \bibinfo
  {author} {\bibfnamefont {R.-M.}\ \bibnamefont {Ma}}, \bibinfo {author}
  {\bibfnamefont {C.}~\bibnamefont {Gladden}}, \bibinfo {author} {\bibfnamefont
  {L.}~\bibnamefont {Dai}}, \bibinfo {author} {\bibfnamefont {G.}~\bibnamefont
  {Bartal}}, \ and\ \bibinfo {author} {\bibfnamefont {X.}~\bibnamefont
  {Zhang}},\ }\href {\doibase 10.1038/nature08364} {\bibfield  {journal}
  {\bibinfo  {journal} {Nature}\ }\textbf {\bibinfo {volume} {461}},\ \bibinfo
  {pages} {629} (\bibinfo {year} {2009})}\BibitemShut {NoStop}%
\bibitem [{\citenamefont {Nomura}\ \emph {et~al.}(2010)\citenamefont {Nomura},
  \citenamefont {Kumagai}, \citenamefont {Iwamoto}, \citenamefont {Ota},\ and\
  \citenamefont {Arakawa}}]{Nomura2010}%
  \BibitemOpen
  \bibfield  {author} {\bibinfo {author} {\bibfnamefont {M.}~\bibnamefont
  {Nomura}}, \bibinfo {author} {\bibfnamefont {N.}~\bibnamefont {Kumagai}},
  \bibinfo {author} {\bibfnamefont {S.}~\bibnamefont {Iwamoto}}, \bibinfo
  {author} {\bibfnamefont {Y.}~\bibnamefont {Ota}}, \ and\ \bibinfo {author}
  {\bibfnamefont {Y.}~\bibnamefont {Arakawa}},\ }\href {\doibase
  10.1038/nphys1518} {\bibfield  {journal} {\bibinfo  {journal} {Nature
  Physics}\ }\textbf {\bibinfo {volume} {6}},\ \bibinfo {pages} {279} (\bibinfo
  {year} {2010})}\BibitemShut {NoStop}%
\bibitem [{\citenamefont {Nezhad}\ \emph {et~al.}(2010)\citenamefont {Nezhad},
  \citenamefont {Simic},\ and\ \citenamefont {Bondarenko}}]{Nezhad2010}%
  \BibitemOpen
  \bibfield  {author} {\bibinfo {author} {\bibfnamefont {M.}~\bibnamefont
  {Nezhad}}, \bibinfo {author} {\bibfnamefont {A.}~\bibnamefont {Simic}}, \
  and\ \bibinfo {author} {\bibfnamefont {O.}~\bibnamefont {Bondarenko}},\
  }\href {\doibase 10.1038/NPHOTON.2010.88} {\bibfield  {journal} {\bibinfo
  {journal} {Nature \ldots}\ }\textbf {\bibinfo {volume} {4}},\ \bibinfo
  {pages} {395} (\bibinfo {year} {2010})}\BibitemShut {NoStop}%
\bibitem [{\citenamefont {Ma}\ \emph {et~al.}(2010)\citenamefont {Ma},
  \citenamefont {Oulton}, \citenamefont {Sorger}, \citenamefont {Bartal},\ and\
  \citenamefont {Zhang}}]{Ma2010}%
  \BibitemOpen
  \bibfield  {author} {\bibinfo {author} {\bibfnamefont {R.-M.}\ \bibnamefont
  {Ma}}, \bibinfo {author} {\bibfnamefont {R.~F.}\ \bibnamefont {Oulton}},
  \bibinfo {author} {\bibfnamefont {V.~J.}\ \bibnamefont {Sorger}}, \bibinfo
  {author} {\bibfnamefont {G.}~\bibnamefont {Bartal}}, \ and\ \bibinfo {author}
  {\bibfnamefont {X.}~\bibnamefont {Zhang}},\ }\href {\doibase
  10.1038/nmat2919} {\bibfield  {journal} {\bibinfo  {journal} {Nature
  Materials}\ }\textbf {\bibinfo {volume} {10}},\ \bibinfo {pages} {110}
  (\bibinfo {year} {2010})}\BibitemShut {NoStop}%
\bibitem [{\citenamefont {Wu}\ \emph {et~al.}(2011)\citenamefont {Wu},
  \citenamefont {Kuo}, \citenamefont {Wang}, \citenamefont {He}, \citenamefont
  {Lin}, \citenamefont {Ahn},\ and\ \citenamefont {Gwo}}]{Wu2011}%
  \BibitemOpen
  \bibfield  {author} {\bibinfo {author} {\bibfnamefont {C.-Y.}\ \bibnamefont
  {Wu}}, \bibinfo {author} {\bibfnamefont {C.-T.}\ \bibnamefont {Kuo}},
  \bibinfo {author} {\bibfnamefont {C.-Y.}\ \bibnamefont {Wang}}, \bibinfo
  {author} {\bibfnamefont {C.-L.}\ \bibnamefont {He}}, \bibinfo {author}
  {\bibfnamefont {M.-H.}\ \bibnamefont {Lin}}, \bibinfo {author} {\bibfnamefont
  {H.}~\bibnamefont {Ahn}}, \ and\ \bibinfo {author} {\bibfnamefont
  {S.}~\bibnamefont {Gwo}},\ }\href {\doibase 10.1021/nl2022477} {\bibfield
  {journal} {\bibinfo  {journal} {Nano letters}\ }\textbf {\bibinfo {volume}
  {11}},\ \bibinfo {pages} {4256} (\bibinfo {year} {2011})}\BibitemShut
  {NoStop}%
\bibitem [{\citenamefont {Lu}\ \emph {et~al.}(2012)\citenamefont {Lu},
  \citenamefont {Kim}, \citenamefont {Chen}, \citenamefont {Wu}, \citenamefont
  {Dabidian}, \citenamefont {Sanders}, \citenamefont {Wang}, \citenamefont
  {Lu}, \citenamefont {Li}, \citenamefont {Qiu}, \citenamefont {Chang},
  \citenamefont {Chen}, \citenamefont {Shvets}, \citenamefont {Shih},\ and\
  \citenamefont {Gwo}}]{Lu2012}%
  \BibitemOpen
  \bibfield  {author} {\bibinfo {author} {\bibfnamefont {Y.-J.}\ \bibnamefont
  {Lu}}, \bibinfo {author} {\bibfnamefont {J.}~\bibnamefont {Kim}}, \bibinfo
  {author} {\bibfnamefont {H.-Y.}\ \bibnamefont {Chen}}, \bibinfo {author}
  {\bibfnamefont {C.}~\bibnamefont {Wu}}, \bibinfo {author} {\bibfnamefont
  {N.}~\bibnamefont {Dabidian}}, \bibinfo {author} {\bibfnamefont {C.~E.}\
  \bibnamefont {Sanders}}, \bibinfo {author} {\bibfnamefont {C.-Y.}\
  \bibnamefont {Wang}}, \bibinfo {author} {\bibfnamefont {M.-Y.}\ \bibnamefont
  {Lu}}, \bibinfo {author} {\bibfnamefont {B.-H.}\ \bibnamefont {Li}}, \bibinfo
  {author} {\bibfnamefont {X.}~\bibnamefont {Qiu}}, \bibinfo {author}
  {\bibfnamefont {W.-H.}\ \bibnamefont {Chang}}, \bibinfo {author}
  {\bibfnamefont {L.-J.}\ \bibnamefont {Chen}}, \bibinfo {author}
  {\bibfnamefont {G.}~\bibnamefont {Shvets}}, \bibinfo {author} {\bibfnamefont
  {C.-K.}\ \bibnamefont {Shih}}, \ and\ \bibinfo {author} {\bibfnamefont
  {S.}~\bibnamefont {Gwo}},\ }\href {\doibase 10.1126/science.1223504}
  {\bibfield  {journal} {\bibinfo  {journal} {Science (New York, N.Y.)}\
  }\textbf {\bibinfo {volume} {337}},\ \bibinfo {pages} {450} (\bibinfo {year}
  {2012})}\BibitemShut {NoStop}%
\bibitem [{\citenamefont {Suh}\ \emph {et~al.}(2012)\citenamefont {Suh},
  \citenamefont {Kim}, \citenamefont {Zhou}, \citenamefont {Huntington},
  \citenamefont {Co}, \citenamefont {Wasielewski},\ and\ \citenamefont
  {Odom}}]{Suh2012}%
  \BibitemOpen
  \bibfield  {author} {\bibinfo {author} {\bibfnamefont {J.~Y.}\ \bibnamefont
  {Suh}}, \bibinfo {author} {\bibfnamefont {C.~H.}\ \bibnamefont {Kim}},
  \bibinfo {author} {\bibfnamefont {W.}~\bibnamefont {Zhou}}, \bibinfo {author}
  {\bibfnamefont {M.~D.}\ \bibnamefont {Huntington}}, \bibinfo {author}
  {\bibfnamefont {D.~T.}\ \bibnamefont {Co}}, \bibinfo {author} {\bibfnamefont
  {M.~R.}\ \bibnamefont {Wasielewski}}, \ and\ \bibinfo {author} {\bibfnamefont
  {T.~W.}\ \bibnamefont {Odom}},\ }\href {\doibase 10.1021/nl303086r}
  {\bibfield  {journal} {\bibinfo  {journal} {Nano letters}\ }\textbf {\bibinfo
  {volume} {12}},\ \bibinfo {pages} {5769} (\bibinfo {year}
  {2012})}\BibitemShut {NoStop}%
\bibitem [{\citenamefont {Bogdanov}\ and\ \citenamefont
  {Fedorov}(2013)}]{Bogdanov2013}%
  \BibitemOpen
  \bibfield  {author} {\bibinfo {author} {\bibfnamefont {A.}~\bibnamefont
  {Bogdanov}}\ and\ \bibinfo {author} {\bibfnamefont {I.}~\bibnamefont
  {Fedorov}},\ }\href {http://arxiv.org/abs/1301.3036} {\bibfield  {journal}
  {\bibinfo  {journal} {arXiv preprint arXiv: \ldots}\ ,\ \bibinfo {pages} {1}}
  (\bibinfo {year} {2013})}\BibitemShut {NoStop}%
\bibitem [{\citenamefont {Protsenko}\ \emph {et~al.}(2005)\citenamefont
  {Protsenko}, \citenamefont {Uskov}, \citenamefont {Zaimidoroga},
  \citenamefont {Samoilov},\ and\ \citenamefont {O’Reilly}}]{Protsenko2005}%
  \BibitemOpen
  \bibfield  {author} {\bibinfo {author} {\bibfnamefont {I.}~\bibnamefont
  {Protsenko}}, \bibinfo {author} {\bibfnamefont {A.}~\bibnamefont {Uskov}},
  \bibinfo {author} {\bibfnamefont {O.}~\bibnamefont {Zaimidoroga}}, \bibinfo
  {author} {\bibfnamefont {V.}~\bibnamefont {Samoilov}}, \ and\ \bibinfo
  {author} {\bibfnamefont {E.}~\bibnamefont {O’Reilly}},\ }\href {\doibase
  10.1103/PhysRevA.71.063812} {\bibfield  {journal} {\bibinfo  {journal}
  {Physical Review A}\ }\textbf {\bibinfo {volume} {71}},\ \bibinfo {pages} {1}
  (\bibinfo {year} {2005})}\BibitemShut {NoStop}%
\bibitem [{\citenamefont {Stockman}(2010)}]{Stockman2010}%
  \BibitemOpen
  \bibfield  {author} {\bibinfo {author} {\bibfnamefont {M.~I.}\ \bibnamefont
  {Stockman}},\ }\href {\doibase 10.1088/2040-8978/12/2/024004} {\bibfield
  {journal} {\bibinfo  {journal} {Journal of Optics}\ }\textbf {\bibinfo
  {volume} {12}},\ \bibinfo {pages} {024004} (\bibinfo {year}
  {2010})}\BibitemShut {NoStop}%
\bibitem [{\citenamefont {Andrianov}\ \emph {et~al.}(2011)\citenamefont
  {Andrianov}, \citenamefont {Pukhov}, \citenamefont {Dorofeenko},
  \citenamefont {Vinogradov},\ and\ \citenamefont {Lisyansky}}]{Andrianov2011}%
  \BibitemOpen
  \bibfield  {author} {\bibinfo {author} {\bibfnamefont {E.~S.}\ \bibnamefont
  {Andrianov}}, \bibinfo {author} {\bibfnamefont {a.~a.}\ \bibnamefont
  {Pukhov}}, \bibinfo {author} {\bibfnamefont {a.~V.}\ \bibnamefont
  {Dorofeenko}}, \bibinfo {author} {\bibfnamefont {a.~P.}\ \bibnamefont
  {Vinogradov}}, \ and\ \bibinfo {author} {\bibfnamefont {a.~a.}\ \bibnamefont
  {Lisyansky}},\ }\href {\doibase 10.1134/S1064226911120151} {\bibfield
  {journal} {\bibinfo  {journal} {Journal of Communications Technology and
  Electronics}\ }\textbf {\bibinfo {volume} {56}},\ \bibinfo {pages} {1471}
  (\bibinfo {year} {2011})}\BibitemShut {NoStop}%
\bibitem [{\citenamefont {Sarychev}\ and\ \citenamefont
  {Tartakovsky}(2007)}]{Sarychev2007}%
  \BibitemOpen
  \bibfield  {author} {\bibinfo {author} {\bibfnamefont {A.}~\bibnamefont
  {Sarychev}}\ and\ \bibinfo {author} {\bibfnamefont {G.}~\bibnamefont
  {Tartakovsky}},\ }\href {\doibase 10.1103/PhysRevB.75.085436} {\bibfield
  {journal} {\bibinfo  {journal} {Physical Review B}\ }\textbf {\bibinfo
  {volume} {75}},\ \bibinfo {pages} {1} (\bibinfo {year} {2007})}\BibitemShut
  {NoStop}%
\bibitem [{\citenamefont {Rice}\ and\ \citenamefont
  {Carmichael}(1994)}]{Rice1994}%
  \BibitemOpen
  \bibfield  {author} {\bibinfo {author} {\bibfnamefont {P.}~\bibnamefont
  {Rice}}\ and\ \bibinfo {author} {\bibfnamefont {H.}~\bibnamefont
  {Carmichael}},\ }\href {http://pra.aps.org/abstract/PRA/v50/i5/p4318\_1}
  {\bibfield  {journal} {\bibinfo  {journal} {Physical Review A}\ }\textbf
  {\bibinfo {volume} {50}},\ \bibinfo {pages} {4318} (\bibinfo {year}
  {1994})}\BibitemShut {NoStop}%
\bibitem [{\citenamefont {Mu}\ and\ \citenamefont {Savage}(1992)}]{Mu1992}%
  \BibitemOpen
  \bibfield  {author} {\bibinfo {author} {\bibfnamefont {Y.}~\bibnamefont
  {Mu}}\ and\ \bibinfo {author} {\bibfnamefont {C.}~\bibnamefont {Savage}},\
  }\href {http://pra.aps.org/abstract/PRA/v46/i9/p5944\_1} {\bibfield
  {journal} {\bibinfo  {journal} {Physical Review A}\ }\textbf {\bibinfo
  {volume} {46}} (\bibinfo {year} {1992})}\BibitemShut {NoStop}%
\bibitem [{\citenamefont {Benson}\ and\ \citenamefont
  {Yamamoto}(1999)}]{Benson1999}%
  \BibitemOpen
  \bibfield  {author} {\bibinfo {author} {\bibfnamefont {O.}~\bibnamefont
  {Benson}}\ and\ \bibinfo {author} {\bibfnamefont {Y.}~\bibnamefont
  {Yamamoto}},\ }\href {\doibase 10.1103/PhysRevA.59.4756} {\bibfield
  {journal} {\bibinfo  {journal} {Physical Review A}\ }\textbf {\bibinfo
  {volume} {59}},\ \bibinfo {pages} {4756} (\bibinfo {year}
  {1999})}\BibitemShut {NoStop}%
\bibitem [{\citenamefont {Protsenko}\ \emph {et~al.}(1999)\citenamefont
  {Protsenko}, \citenamefont {Domokos}, \citenamefont {Lef\`{e}vre-Seguin},
  \citenamefont {Hare}, \citenamefont {Raimond},\ and\ \citenamefont
  {Davidovich}}]{Protsenko1999}%
  \BibitemOpen
  \bibfield  {author} {\bibinfo {author} {\bibfnamefont {I.}~\bibnamefont
  {Protsenko}}, \bibinfo {author} {\bibfnamefont {P.}~\bibnamefont {Domokos}},
  \bibinfo {author} {\bibfnamefont {V.}~\bibnamefont {Lef\`{e}vre-Seguin}},
  \bibinfo {author} {\bibfnamefont {J.}~\bibnamefont {Hare}}, \bibinfo {author}
  {\bibfnamefont {J.}~\bibnamefont {Raimond}}, \ and\ \bibinfo {author}
  {\bibfnamefont {L.}~\bibnamefont {Davidovich}},\ }\href {\doibase
  10.1103/PhysRevA.59.1667} {\bibfield  {journal} {\bibinfo  {journal}
  {Physical Review A}\ }\textbf {\bibinfo {volume} {59}},\ \bibinfo {pages}
  {1667} (\bibinfo {year} {1999})}\BibitemShut {NoStop}%
\bibitem [{\citenamefont {Hill}\ \emph {et~al.}(2007)\citenamefont {Hill},
  \citenamefont {Oei}, \citenamefont {Smalbrugge}, \citenamefont {Zhu},
  \citenamefont {de~Vries}, \citenamefont {van Veldhoven}, \citenamefont {van
  Otten}, \citenamefont {Eijkemans}, \citenamefont {Turkiewicz}, \citenamefont
  {de~Waardt}, \citenamefont {Geluk}, \citenamefont {Kwon}, \citenamefont
  {Lee}, \citenamefont {N\"{o}tzel},\ and\ \citenamefont {Smit}}]{Hill2007}%
  \BibitemOpen
  \bibfield  {author} {\bibinfo {author} {\bibfnamefont {M.~T.}\ \bibnamefont
  {Hill}}, \bibinfo {author} {\bibfnamefont {Y.-S.}\ \bibnamefont {Oei}},
  \bibinfo {author} {\bibfnamefont {B.}~\bibnamefont {Smalbrugge}}, \bibinfo
  {author} {\bibfnamefont {Y.}~\bibnamefont {Zhu}}, \bibinfo {author}
  {\bibfnamefont {T.}~\bibnamefont {de~Vries}}, \bibinfo {author}
  {\bibfnamefont {P.~J.}\ \bibnamefont {van Veldhoven}}, \bibinfo {author}
  {\bibfnamefont {F.~W.~M.}\ \bibnamefont {van Otten}}, \bibinfo {author}
  {\bibfnamefont {T.~J.}\ \bibnamefont {Eijkemans}}, \bibinfo {author}
  {\bibfnamefont {J.~P.}\ \bibnamefont {Turkiewicz}}, \bibinfo {author}
  {\bibfnamefont {H.}~\bibnamefont {de~Waardt}}, \bibinfo {author}
  {\bibfnamefont {E.~J.}\ \bibnamefont {Geluk}}, \bibinfo {author}
  {\bibfnamefont {S.-H.}\ \bibnamefont {Kwon}}, \bibinfo {author}
  {\bibfnamefont {Y.-H.}\ \bibnamefont {Lee}}, \bibinfo {author} {\bibfnamefont
  {R.}~\bibnamefont {N\"{o}tzel}}, \ and\ \bibinfo {author} {\bibfnamefont
  {M.~K.}\ \bibnamefont {Smit}},\ }\href {\doibase 10.1038/nphoton.2007.171}
  {\bibfield  {journal} {\bibinfo  {journal} {Nature Photonics}\ }\textbf
  {\bibinfo {volume} {1}},\ \bibinfo {pages} {589} (\bibinfo {year}
  {2007})}\BibitemShut {NoStop}%
\bibitem [{\citenamefont {Kang}\ \emph {et~al.}(2011)\citenamefont {Kang},
  \citenamefont {Park},\ and\ \citenamefont {Kwon}}]{Kang2011}%
  \BibitemOpen
  \bibfield  {author} {\bibinfo {author} {\bibfnamefont {J.}~\bibnamefont
  {Kang}}, \bibinfo {author} {\bibfnamefont {H.}~\bibnamefont {Park}}, \ and\
  \bibinfo {author} {\bibfnamefont {S.}~\bibnamefont {Kwon}},\ }\href
  {http://www.opticsinfobase.org/abstract.cfm?URI=oe-19-15-13892} {\bibfield
  {journal} {\bibinfo  {journal} {Optics Express}\ }\textbf {\bibinfo {volume}
  {19}},\ \bibinfo {pages} {1537} (\bibinfo {year} {2011})}\BibitemShut
  {NoStop}%
\bibitem [{\citenamefont {Scully}\ and\ \citenamefont
  {Zubairy}(1997)}]{Scully1997}%
  \BibitemOpen
  \bibfield  {author} {\bibinfo {author} {\bibfnamefont {M.}~\bibnamefont
  {Scully}}\ and\ \bibinfo {author} {\bibfnamefont {M.}~\bibnamefont
  {Zubairy}},\ }\href
  {http://books.google.com/books?hl=en\&lr=\&id=20ISsQCKKmQC\&oi=fnd\&pg=PR19\&dq=Quantum+Optics\&ots=yRTSRMHDst\&sig=W-FT3Cqa6X9yD8yTk7XPcUOp8MM}
  {\emph {\bibinfo {title} {{Quantum optics}}}}\ (\bibinfo {year}
  {1997})\BibitemShut {NoStop}%
\bibitem [{\citenamefont {Meystre}\ and\ \citenamefont
  {Sargent}(1999)}]{Meystre1999}%
  \BibitemOpen
  \bibfield  {author} {\bibinfo {author} {\bibfnamefont {P.}~\bibnamefont
  {Meystre}}\ and\ \bibinfo {author} {\bibfnamefont {M.}~\bibnamefont
  {Sargent}},\ }\href
  {http://books.google.com/books?hl=en\&lr=\&id=dWnIOHloxoEC\&oi=fnd\&pg=PA1\&dq=Elements+of+Quantum+Optics\&ots=FFbD6c0Giy\&sig=tyXfGIe0TAHbtrwz0zfmNBKa8WA}
  {\emph {\bibinfo {title} {{Elements of quantum optics}}}}\ (\bibinfo {year}
  {1999})\BibitemShut {NoStop}%
\bibitem [{\citenamefont {Greenstein}\ and\ \citenamefont
  {Zajonc}(1995)}]{Greenstein1995}%
  \BibitemOpen
  \bibfield  {author} {\bibinfo {author} {\bibfnamefont {G.}~\bibnamefont
  {Greenstein}}\ and\ \bibinfo {author} {\bibfnamefont {A.}~\bibnamefont
  {Zajonc}},\ }\href {http://adsabs.harvard.edu/abs/1995AmJPh..63..743G}
  {\bibfield  {journal} {\bibinfo  {journal} {American Journal of Physics}\ }
  (\bibinfo {year} {1995})}\BibitemShut {NoStop}%
\bibitem [{\citenamefont {Milburn}(2008)}]{Milburn2008}%
  \BibitemOpen
  \bibfield  {author} {\bibinfo {author} {\bibfnamefont {D.~G.~J.}\
  \bibnamefont {Milburn}},\ }\href
  {http://books.google.com/books?hl=en\&lr=\&id=20ISsQCKKmQC\&oi=fnd\&pg=PR19\&dq=Quantum+Optics\&ots=yRTRMSHHms\&sig=dUqnEVpKtlXJqgPCA7lMnslT\_G4}
  {\emph {\bibinfo {title} {{Quantum optics}}}}\ (\bibinfo {year}
  {2008})\BibitemShut {NoStop}%
\bibitem [{\citenamefont {Bj\"{o}rk}\ \emph {et~al.}(1994)\citenamefont
  {Bj\"{o}rk}, \citenamefont {Karlsson},\ and\ \citenamefont
  {Yamamoto}}]{Bjork1994}%
  \BibitemOpen
  \bibfield  {author} {\bibinfo {author} {\bibfnamefont {G.}~\bibnamefont
  {Bj\"{o}rk}}, \bibinfo {author} {\bibfnamefont {A.}~\bibnamefont {Karlsson}},
  \ and\ \bibinfo {author} {\bibfnamefont {Y.}~\bibnamefont {Yamamoto}},\
  }\href {http://pra.aps.org/abstract/PRA/v50/i2/p1675\_1} {\bibfield
  {journal} {\bibinfo  {journal} {Physical Review A}\ }\textbf {\bibinfo
  {volume} {50}},\ \bibinfo {pages} {1675} (\bibinfo {year}
  {1994})}\BibitemShut {NoStop}%
\bibitem [{\citenamefont {Bykov}(1994)}]{Bykov1994}%
  \BibitemOpen
  \bibfield  {author} {\bibinfo {author} {\bibfnamefont {V.}~\bibnamefont
  {Bykov}},\ }\href
  {http://books.google.com/books?hl=en\&lr=\&id=oNZEZVVYBMcC\&oi=fnd\&pg=PA1\&dq=Radiation+of+Atoms+in+a+Resonant+Environment\&ots=JU75nEt4IG\&sig=Me\_fxnvORlcQgspC9yfD-LeIgZo}
  {\emph {\bibinfo {title} {{Radiation of atoms in a resonant environment}}}}\
  (\bibinfo {year} {1994})\BibitemShut {NoStop}%
\bibitem [{\citenamefont {Carmichael}(1999)}]{Carmichael1999}%
  \BibitemOpen
  \bibfield  {author} {\bibinfo {author} {\bibfnamefont {H.}~\bibnamefont
  {Carmichael}},\ }\href
  {http://scholar.google.com/scholar?hl=en\&btnG=Search\&q=intitle:Statistical+Methods+in+Quantum+Optics+1:+Master+Equations+and+Fokker-Planck+Equations\#4}
  {\emph {\bibinfo {title} {{Quantum Statistical Methods in Quantum Optics 1:
  Master Equations and Fokker-Planck Equations}}}}\ (\bibinfo {year}
  {1999})\BibitemShut {NoStop}%
\bibitem [{\citenamefont {Glauber}(2007)}]{Glauber2007}%
  \BibitemOpen
  \bibfield  {author} {\bibinfo {author} {\bibfnamefont {R.}~\bibnamefont
  {Glauber}},\ }\href
  {http://books.google.com/books?hl=en\&lr=\&id=9V3GzE6iqOYC\&oi=fnd\&pg=PR5\&dq=Quantum+Theory+of+Optical+Coherence\&ots=1sP9LNOq29\&sig=eRGnAM7PZbAWRse6sILzfGSOfLE}
  {\emph {\bibinfo {title} {{Quantum theory of optical coherence}}}}\ (\bibinfo
  {year} {2007})\BibitemShut {NoStop}%
\bibitem [{\citenamefont {Gardiner}\ and\ \citenamefont
  {Zoller}(2004)}]{Gardiner2004}%
  \BibitemOpen
  \bibfield  {author} {\bibinfo {author} {\bibfnamefont {C.}~\bibnamefont
  {Gardiner}}\ and\ \bibinfo {author} {\bibfnamefont {P.}~\bibnamefont
  {Zoller}},\ }\href
  {http://books.google.com/books?hl=en\&lr=\&id=a\_xsT8oGhdgC\&oi=fnd\&pg=PA1\&dq=Quantum+Noise.+A+Handbook+of+Markovian+and+Non-Markovian+Quantum+Stochastic+Methods+with+Applications+to+Quantum+Optics\&ots=kYu1xXiZsf\&sig=U5bYMUYdtKmzdA--5G8QEcAqsyE}
  {\emph {\bibinfo {title} {{Quantum noise: a handbook of Markovian and
  non-Markovian quantum stochastic methods with applications to quantum
  optics}}}}\ (\bibinfo {year} {2004})\BibitemShut {NoStop}%
\end{thebibliography}%

\end{document}